\documentclass[11pt]{article}
\usepackage[a4paper,margin=1in]{geometry}
\usepackage[T1]{fontenc}
\usepackage{lmodern}
\usepackage{microtype}
\usepackage{amsmath,amssymb}
\usepackage{booktabs}
\usepackage{tabularx}
\usepackage{graphicx}
\usepackage{xcolor}
\usepackage[numbers,sort&compress]{natbib}
\usepackage[colorlinks=true,linkcolor=black,citecolor=blue!50!black,urlcolor=blue!50!black]{hyperref}
\usepackage{enumitem}
\usepackage{seqsplit}
\newcommand{\hash}[1]{{\footnotesize\texttt{\seqsplit{#1}}}}
\setlist{itemsep=2pt,topsep=4pt}
\newcolumntype{L}{>{\raggedright\arraybackslash}X}

\title{Limits of Per-Group Privacy Gates for Synthetic Tables:\\
Sample-Size Floors, Pooled Masking, and Three Failed Specifications}
\author{Aniket Tapre\\
Fahrenheit Research\\
\texttt{aniket@fahrenheitresearch.si}}
\date{Version 1.1, November 2026}

\begin{document}
\maketitle

\begin{abstract}
A release gate is an automatic pass or fail decision on a synthetic table before the table leaves the data owner. We specified one with attack-based privacy scores, per-group scoring and planted controls with known answers, and tested it on two public tables: Bank Marketing (45,211 rows) and Diabetes 130-US Hospitals (101,766 rows). Three findings stand.
(1)~\textbf{Per-group privacy testing has hard sample-size floors.} At a 1\% false-positive rate and a 5\% leak tolerance, a group needs 49 members before a leak at the tolerance boundary produces at least one attack hit with 95\% probability, and a clean group needs 152 members (with 100 non-members) to pass 95\% of the time. We give the exact binomial construction for both floors and show that attacks with high baseline success, such as attribute inference, need thousands of records per arm.
(2)~\textbf{A pooled privacy score hides a fully exposed rare class.} When every record of a group holding a share $g$ of the training data is exposed, the pooled leak equals $g$, so a pooled check with tolerance $t$ fires only when $g>t$. A planted control exposing every record of a 3\% class was recovered in full by a distance attack (741 of 741 records on Bank Marketing, 1,678 of 1,678 on Diabetes), yet the pooled distance score read 89.83 to 91.05 against a pass mark of 85. The rule held on five splits never used in design, and an independent reimplementation reproduces it on both tables. With per-class leaks $u_k$ the pooled leak is $\sum_k g_k u_k$, so partial leaks are masked too: a release of real majority-class rows with SMOTE-interpolated rare-class rows exposed 16\% of rare-class members to a distance attack and passed the pooled check at every tolerance on three of three seeds. Gaussian copula, CTGAN, TVAE and an MST-style differentially private generator showed no leak above tolerance.
(3)~\textbf{Per-group gating, as we specified it, rejects clean data.} Three successive specifications were tested. The first passed the clean control in 0 of 9 trials, the second left 6 to 14 components unresolved on every split, and the third, frozen and run prospectively, passed 0 of 5 untouched splits. Sample-size floors, an absolute coverage threshold and many components on small groups caused the failures.
Findings 1 and 2 are not new in kind; disparate vulnerability across subgroups and the difficulty of estimating it for small groups are established. Our contribution is a set of worked numbers for the gate setting, a one-line masking rule, and a complete record of three failures, which we offer as a negative result.
\end{abstract}

\paragraph{Status and disclosure.} This is a working paper and has not been peer reviewed. The gate results come from planted control releases with a known answer. Trained generators were tested only on the masking result, on one table, with an independent implementation (Section~\ref{sec:generators}). The author develops Grafta, a synthetic-table product with a release gate, and has a commercial interest in this topic (Section~\ref{sec:disclosure}).

\section{Introduction}
Teams training small models on private tables want a synthetic copy they are free to share. The copy has two jobs: it must stay useful for training, and it must not expose the people in the source table. Most vendors check the second job with distance metrics, and recent work shows those metrics miss real leakage \citep{ganev2025inadequacy,yao2025dcr,ward2025synthmia}.

We set out to build a stricter check: a release gate. The gate runs privacy attacks against a candidate release, scores each population group separately, takes the worst score, and blocks the release when that score falls below a fixed mark. No human override exists. Planted controls with known answers test the gate itself.

The plan did not survive testing. Each of three specifications rejected a clean control. Two results held across every round, and they explain why the gate is hard to build. This paper reports:
\begin{enumerate}
\item the group sizes below which a per-group privacy test has no resolving power, with an exact construction and worked values (Section~\ref{sec:floors});
\item a masking rule for pooled scores, confirmed on five untouched splits across two tables and by an independent reimplementation (Section~\ref{sec:masking});
\item three failed specifications, the cause of each failure, and the audit trail (Section~\ref{sec:failures});
\item practical guidance for anyone releasing synthetic tables today (Section~\ref{sec:guidance}).
\end{enumerate}
The paper makes no claim about any generator, and it does not show a working gate.

\section{Related work}
\paragraph{Distance metrics fail.} \citet{ganev2025inadequacy} attack the similarity tests used by commercial vendors and reconstruct 78\% to 100\% of training outliers. \citet{yao2025dcr} show that releases passing Distance to Closest Record tests remain open to membership inference. \citet{ward2025synthmia} provide 13 attacks in one library and find that no single attack wins across settings, and \citet{ward2025ensemble} show that ensembles of attacks are a more robust choice than any one attack. Our gate follows this advice and scores attacks, not distances.

\paragraph{Risk concentrates in subgroups and outliers.} \citet{kulynych2022disparate} give the theory of disparate vulnerability to membership inference, show that the size of the disparity is hard to estimate when subgroups are small, and supply a statistical framework for the estimate. \citet{stadler2022groundhog} find that synthetic data either fails to stop inference attacks or loses its utility. \citet{vanbreugel2023domias} report higher attack success on uncommon samples. \citet{krco2024averages} report an average-case evaluation missing up to 94\% of high-risk records. \citet{ganev2025smote} show that minority-class oversampling exposes minority records disproportionately. \citet{lostracco2026whobears} show that the most exposed subgroups in synthetic tables shift with the generator, and that aggregate metrics surface neither pattern. \citet{sun2026subgroup} audit subgroup membership in synthetic text and find leakage concentrated in about a tenth of records, and \citet{pera2026sok} find that records with unique signatures remain highly vulnerable under common defences. Findings 1 and 2 of this paper sit inside this line of work. We add exact floors for a gate setting and a rule tying pooled masking to group share.

\paragraph{Attack design.} \citet{carlini2022first} argue for reporting true-positive rate at a low false-positive rate; we use 1\%. \citet{giomi2023anonymeter} score singling out, linkability and inference against a control set of unseen records, and our control-adjusted risk follows their design. Our density-ratio attack follows \citet{vanbreugel2023domias}.

\paragraph{Release audits.} \citet{opoku2026auditlock} lock the intended use, tolerances and audit schedule before evaluation, test negative controls, and derive a pre-audit sample-size rule. Their work is the closest to ours in intent. They report generators passing once fitted on enough data; we report a gate failing on clean data when scored per group. The two results are compatible: their audit is use-specific and pooled, ours is per-group.

\paragraph{Usefulness.} \citet{alaa2022faithful} define sample-level precision and recall for synthetic data. Our coverage measure is a recall of this type.

We did not run a systematic literature review. Prior statements of the masking rule in Section~\ref{sec:masking} may exist, and we would welcome pointers.

\section{The gate under test}\label{sec:gate}
\subsection{Data split}
Each table is split by entity into four parts: 60\% training, 10\% curation oracle, 10\% reference and 20\% holdout. On Diabetes the entity is the patient: the table holds 101,766 encounters from 71,518 patients, and 16,773 patients appear more than once, so splitting by row would place one patient on both sides of an attack. Each table also gets a rare-class arm in which the positive class is downsampled to 3\% of training rows. Bank Marketing gives 24,694 training rows with 741 positives and a holdout of 9,043 rows. Diabetes gives 55,923 training rows with 1,678 positives and a holdout of 20,353 rows.

\subsection{Privacy score}
Each attack runs at a false-positive rate $\alpha=0.01$, with the threshold calibrated on reference non-members. The leak is
\begin{equation}
u=\frac{\max(0,\mathrm{TPR}-\alpha)}{1-\alpha}.
\end{equation}
With tolerance $t$, the component score is $s=\max(0,\,100-15u/t)$, so a leak equal to the tolerance scores 85, the pass mark. The privacy score $P$ is the lowest component score across attacks and groups. We tested tolerances of 0.02, 0.03 and 0.05. Groups are the pooled table, each outcome class, and a vulnerable-record group sized by count. The decision uses a one-sided 95\% upper bound on the leak, not the point estimate. A component returns one of five labels: \texttt{pass}, \texttt{fail\_exposure}, \texttt{unresolvable\_sample\_size}, \texttt{not\_assessed} or \texttt{invalid}. Table~\ref{tab:tiers} lists the attacks.

\begin{table}[h]
\centering\small
\caption{Attack tiers in the third specification. The first two specifications scored every attack in every group.}\label{tab:tiers}
\begin{tabularx}{\linewidth}{lLL}
\toprule
Tier & Attacks & Where scored\\
\midrule
1 & Exact and near-copy checks; distance-based membership inference; density-ratio membership inference & Every group\\
2 & Linkability; attribute inference; both adjusted against a control set of unseen records & Pooled table and groups of adequate size\\
\bottomrule
\end{tabularx}
\end{table}

\subsection{Usefulness score}
$T_1$ is the train-on-synthetic, test-on-real ratio and $Q_1$ is query agreement. The usefulness score is $U=100\cdot\min(T_1,Q_1)$ and passes at 90. Two floors apply: a worst-group transfer ratio $T_2\geq 0.80$ and nearest-neighbour coverage $K_1\geq 0.80$.

\subsection{Planted controls}
A gate deserves trust only when it gives the right answer on releases with a known answer. Table~\ref{tab:controls} lists the five controls. The acceptance test is simple: Clean passes and the other four fail.

\begin{table}[h]
\centering\small
\caption{Planted controls.}\label{tab:controls}
\begin{tabularx}{\linewidth}{lLl}
\toprule
Control & Construction & Correct decision\\
\midrule
Clean & Real records from the curation oracle, with no link to the training split & Pass\\
Copy & Training records released unchanged & Fail\\
Leaky & A share of training records exposed & Fail\\
Rare-leak & Every rare-class training record exposed, remainder clean & Fail\\
Shuffled & Dependence between columns destroyed & Fail on usefulness\\
\bottomrule
\end{tabularx}
\end{table}

\section{Finding 1: sample-size floors}\label{sec:floors}
A per-group test needs enough records in the group. Below a floor the test returns noise, and a strict gate turns noise into rejections.

\subsection{Detection floor}
Take a group whose true leak sits at the tolerance boundary. The attack hits each member with probability $\alpha+t(1-\alpha)$. The chance of zero hits in $n$ members falls below 5\% only when
\begin{equation}
n\;\geq\;\frac{\ln 0.05}{\ln\bigl(1-\alpha-t(1-\alpha)\bigr)}.
\end{equation}
Below this size, a group leaking at the boundary shows no hits at all more than one time in twenty.

\subsection{Adequacy floor}
\begin{figure}[h]
\centering
\IfFileExists{fig_floors.pdf}{\includegraphics[width=0.72\linewidth]{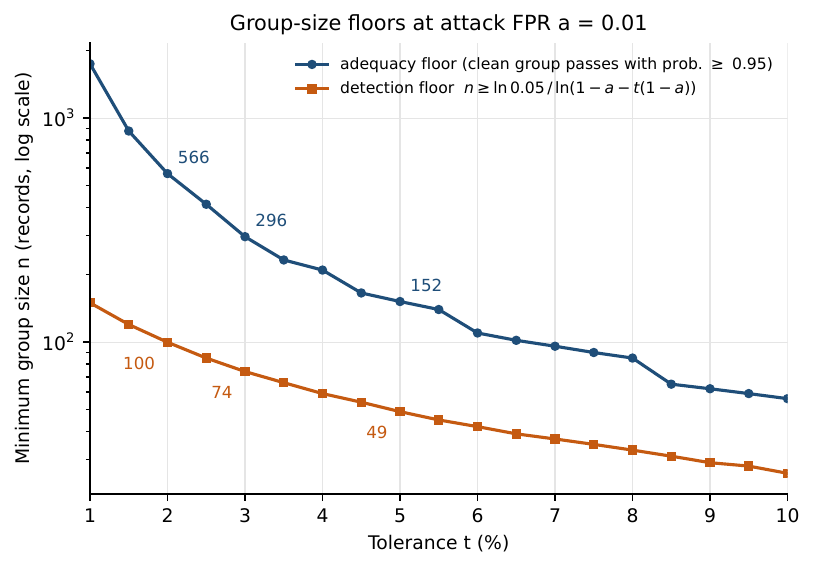}}{\fbox{fig\_floors.pdf}}
\caption{Detection and adequacy floors as a function of the tolerance $t$ at $\alpha=0.01$.}\label{fig:floors}
\end{figure}

Detection is the weaker requirement; a gate also needs a clean group to pass. Let a clean group of $n$ members have true $\mathrm{TPR}=\alpha$, so that the hit count is $K\sim\mathrm{Binomial}(n,\alpha)$. The group passes when the one-sided 95\% Clopper-Pearson upper bound on its TPR, $\mathrm{B}^{-1}(0.95;\,K+1,\,n-K)$, does not exceed $\alpha+t(1-\alpha)$, the largest true TPR the tolerance accepts. Because $K$ is discrete, the pass probability is not monotone in $n$. We define the adequacy floor as the smallest $n$ from which the pass probability stays at or above 95\% for every larger $n$. We require 100 non-members for threshold calibration in addition. Table~\ref{tab:floors} gives the values; Figure~\ref{fig:floors} shows both floors across tolerances. Both columns are exact computations and are reproduced by the public script described in Appendix~\ref{app:replication}.

\begin{table}[h]
\centering\small
\caption{Sample-size floors at $\alpha=0.01$ and 95\% confidence.}\label{tab:floors}
\begin{tabular}{cccc}
\toprule
Tolerance $t$ & Detection floor (members) & Adequacy floor (members) & Largest true TPR accepted\\
\midrule
0.02 & 100 & 566 & 2.98\%\\
0.03 & 74 & 296 & 3.97\%\\
0.05 & 49 & 152 & 5.95\%\\
\bottomrule
\end{tabular}
\end{table}

\subsection{Attacks with a high baseline need far more}
Linkability and attribute inference compare a success rate on training records against the rate $p$ on unseen control records. Following \citet{giomi2023anonymeter}, the control-adjusted risk is $(r_{\text{train}}-p)/(1-p)$. Under the null of no leak, its standard error is approximately $\sqrt{p(1-p)(1/n+1/h)}/(1-p)=\sqrt{\tfrac{p}{1-p}(1/n+1/h)}$ for $n$ training and $h$ control records. A clean component passes 95\% of the time when the estimate plus one-sided margin stays below tolerance in 95\% of draws, which gives the planning condition
\begin{equation}
2\times 1.645\,\sqrt{\frac{p}{1-p}\Bigl(\frac{1}{n}+\frac{1}{h}\Bigr)}\;\leq\;t.
\end{equation}
The measured baseline rates ran from 38\% to 53\%, because an attacker guesses a common attribute value correctly without any leak. At $t=0.05$ with equal arms, the condition needs about 5,307 records per arm at a 38\% baseline and about 9,763 at 53\%. No rare class in our tables reaches those counts, so per-group scoring of these attacks is out of reach for small groups at any tolerance we tested.

\subsection{What the floor excludes}
Two more tables entered data preparation and left it. German Credit kept 13 rare-class records in the 3\% arm and SECOM kept 27, both below the detection floor of 49. With zero hits, the highest attainable privacy score for those groups is 40.66 and 71.21, below the pass mark of 85. A per-group gate has no valid decision on tables of this size.

\section{Finding 2: a pooled score hides a fully exposed rare class}\label{sec:masking}
\subsection{The rule}
\begin{figure}[h]
\centering
\IfFileExists{fig_masking.pdf}{\includegraphics[width=0.72\linewidth]{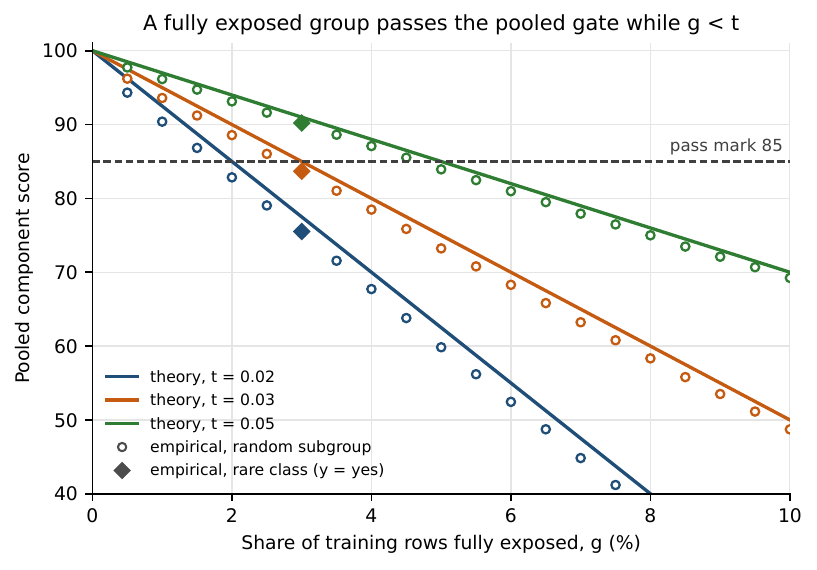}}{\fbox{fig\_masking.pdf}}
\caption{Pooled privacy score against the share $g$ of training records in a fully exposed group. Lines show $100-15g/t$; points are measured with the independent reimplementation (Appendix~\ref{app:replication}). The dashed line is the pass mark of 85.}\label{fig:masking}
\end{figure}

Suppose a group holds a share $g$ of the training records and every record in it is exposed. An attack hits the whole group and hits other records at the false-positive rate, so the pooled true-positive rate is $\alpha+g(1-\alpha)$ and
\begin{equation}
u_{\text{pooled}}=g.
\end{equation}
The pooled leak equals the group share, and a pooled check with tolerance $t$ fires only when $g>t$. A fully exposed class of 3\% scores $100-15\times0.03/0.05=91$ at a tolerance of 0.05. The release passes with six points to spare while 3\% of the people in the table are fully exposed. Figure~\ref{fig:masking} plots the rule against measured points.

\subsection{The evidence}
The Rare-leak control exposes every rare-class training record and nothing else. The distance attack recovered every exposed record: 741 of 741 on Bank Marketing and 1,678 of 1,678 on Diabetes. The pooled distance scores at a tolerance of 0.05 were 91.05, 90.21 and 90.02 on the first three Bank Marketing splits. On five further splits, Bank Marketing 1747 and 1753 and Diabetes 1729, 1733 and 1741, none of which was used to design the specification, the scores ran from 89.83 to 90.42 against a predicted 91. Table~\ref{tab:pooled} counts the splits on which each pooled check passed the Rare-leak control.

\begin{table}[h]
\centering\small
\caption{Pooled checks that passed the Rare-leak control on five untouched splits.}\label{tab:pooled}
\begin{tabular}{lccc}
\toprule
Pooled check & $t=0.02$ & $t=0.03$ & $t=0.05$\\
\midrule
Distance-based membership & 0 of 5 & 0 of 5 & 5 of 5\\
Density-ratio membership & 5 of 5 & 5 of 5 & 5 of 5\\
Linkability & 5 of 5 & 5 of 5 & 5 of 5\\
All attacks combined & 0 of 5 & 0 of 5 & 3 of 5\\
\bottomrule
\end{tabular}
\end{table}

The distance row matches the rule. The upper bound on the pooled leak ran from 0.032 to 0.034: above 0.02 and 0.03 and below 0.05. The share is 3\%, so the check fires at the two tighter tolerances and stays silent at the loose one. Density ratio and linkability missed the exposed class at every tolerance when pooled, so attack choice matters as much as tolerance. The full gate with per-group scoring rejected the Rare-leak control on 5 of 5 splits at every tolerance.

\subsection{A caution on the combined row}
On the first three splits the combined pooled gate rejected the Rare-leak control at 0.05 with scores of 76.81, 83.68 and 78.36. Attribute inference set those scores, not the exposed class, and the same component also rejected the clean control. A rejection for the wrong reason is not detection.

\subsection{Partial leaks and trained generators}\label{sec:generators}
Full exposure is the extreme case. With one shared threshold, the pooled true-positive rate is the share-weighted mean of the group rates, so
\begin{equation}
u_{\text{pooled}}=\sum_k g_k\,u_k ,
\end{equation}
and a group with share $g$ and leak $u$ is masked whenever $g\,u\le t$ while $u>t$. A group of 3\% cannot fail a pooled check at $t=0.05$ whatever it leaks, since that would need $u>t/g\approx1.67$. In our runs the identity held exactly with a shared threshold and to within 0.0043 with clipping or class-calibrated thresholds.

\begin{figure}[h]
\centering
\IfFileExists{fig_generators.pdf}{\includegraphics[width=0.9\linewidth]{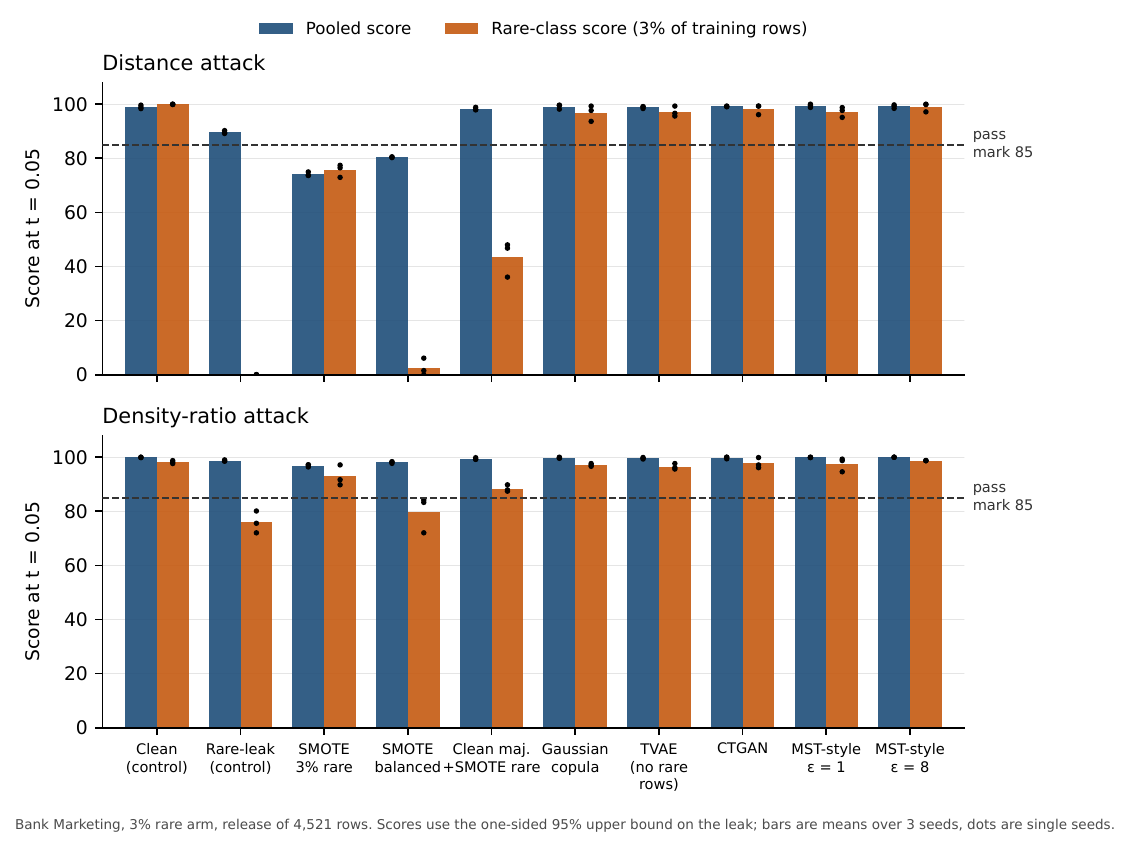}}{\fbox{fig\_generators.pdf}}
\caption{Pooled and rare-class scores at $t=0.05$ for the controls and the trained generators of Table~\ref{tab:generators}, under the distance and density-ratio attacks. Bars are means over three seeds; dots are single seeds. The dashed line is the pass mark of 85.}\label{fig:generators}
\end{figure}

To see whether trained generators produce this pattern, the independent implementation (Appendix~\ref{app:replication}) generated releases of 4,521 rows on the Bank Marketing 3\% arm for seeds 1729, 1733 and 1741 (Table~\ref{tab:generators}, Figure~\ref{fig:generators}). The generators were a Gaussian copula, CTGAN (30 epochs), TVAE (150 epochs), an MST-style marginal generator at $\varepsilon=1$ and $\varepsilon=8$ written for this study without Private-PGM, and SMOTE interpolation of training records. Group thresholds were calibrated on holdout records of the same class.

\begin{table}[h]
\centering\footnotesize\setlength{\tabcolsep}{4pt}
\caption{Trained generators on Bank Marketing (distance attack, three seeds). Leaks are means; scores use the 95\% upper bound at $t=0.05$. Masked counts seeds on which the pooled score passed and the rare-class score failed, at $t=0.02$, $0.03$ and $0.05$.}\label{tab:generators}\smallskip
\begin{tabular}{lccccc}
\toprule
Release & Pooled $u$ & Rare $u$ & Pooled score & Rare score & Masked\\
\midrule
Gaussian copula & 0.002 & 0.003 & 99.1 & 96.9 & 1, 0, 0\\
CTGAN & 0.002 & 0.001 & 99.2 & 98.2 & 0, 0, 0\\
TVAE & 0.003 & 0.002 & 98.9 & 97.2 & 0, 0, 0\\
MST-style, $\varepsilon=1$ & 0.001 & 0.002 & 99.5 & 97.2 & 0, 0, 0\\
MST-style, $\varepsilon=8$ & 0.002 & 0.000 & 99.2 & 99.0 & 0, 0, 0\\
SMOTE, training class mix & 0.083 & 0.063 & 74.1 & 75.6 & 0, 0, 0\\
SMOTE, balanced & 0.063 & 0.307 & 80.4 & 2.5 & 0, 0, 0\\
Real majority, SMOTE rare & 0.004 & 0.163 & 98.3 & 43.6 & 3, 3, 3\\
\bottomrule
\end{tabular}
\end{table}

Three points follow. First, masking is not a property of the mainstream generators: none of them leaked above tolerance, pooled or in the rare class, on this table. Second, a realistic release does produce it. Real majority-class rows combined with SMOTE-interpolated rare-class rows, a common way to rebalance a table before sharing it, exposed between 15\% and 19\% of rare-class members and passed the pooled check at every tolerance on every seed; the density-ratio attack was masked in the same way at $t=0.02$ and $0.03$. When SMOTE was applied to the whole table, the leak was spread across both classes and the pooled check caught it. Third, TVAE produced 0 to 2 rare-class rows per release. That is a usefulness failure that no privacy component can see.

\subsection{Consequence}
A pooled privacy score is blind to any fully exposed group smaller than the tolerance, and more generally to any group whose share times leak stays under it. A tolerance of 5\% leaves every group under 5\% of the table unprotected by the pooled number. Scoring groups separately removes the blind spot and runs straight into the floors of Section~\ref{sec:floors}.

\section{Finding 3: three specifications, three failures on clean data}\label{sec:failures}
Every specification rejected the leaking controls: Copy, Leaky, Rare-leak and Shuffled failed in every round. Rejecting bad releases is the easy half. Passing a good one is the hard half, and no specification managed it (Table~\ref{tab:specs}).

\begin{table}[h]
\centering\small
\caption{Clean-control outcomes across the three specifications.}\label{tab:specs}
\begin{tabularx}{\linewidth}{llLL}
\toprule
Specification & Test data & Clean control result & Cause\\
\midrule
First & Bank Marketing, 3 splits, 3 tolerances & Passed 0 of 9 & Conservative bounds on small groups and on high-baseline attacks\\
Second (amended) & Same 3 splits, replayed & 6 to 14 components unresolved per split & Groups below the adequacy floor\\
Third (two tiers) & 5 untouched splits, 2 tables & Passed 0 of 5 at every tolerance & Coverage floor, residual groups, multiple testing\\
\bottomrule
\end{tabularx}
\end{table}

\subsection{First specification}
The clean control failed all nine trials. The clearest case is attribute inference on an age band: the observed excess over the control baseline was 0.000 and the upper bound was 0.096, nearly twice the tolerance of 0.05. The release showed no leak and the gate still blocked it, for the reason given in Section~\ref{sec:floors}.

\subsection{Second specification}
We made eight amendments after seeing the first results, so this round is retrospective and carries no confirmatory weight. A tighter one-sided bound on the difference cut three example bounds from 0.0956, 0.0425 and 0.0563 to 0.0573, 0.0024 and 0.0169. Clean components then split as shown in Table~\ref{tab:second}.

\begin{table}[h]
\centering\small
\caption{Clean components under the second specification, per split (seeds 1729, 1733, 1741).}\label{tab:second}
\begin{tabular}{cccc}
\toprule
Tolerance & Passed & False alarms & Unresolved (each split)\\
\midrule
0.02 & 33, 34, 33 & 1, 0, 0 & 14\\
0.03 & 34, 34, 34 & 0, 0, 0 & 14\\
0.05 & 42, 42, 43 & 1, 1, 0 & 6\\
\bottomrule
\end{tabular}
\end{table}

False alarms nearly vanished; the unresolved components did not. A gate with no override has no way to release a table carrying 6 to 14 open components. One amendment was wrong outright: we proposed class-mix reweighting to repair coverage, and the replay did not support it.

\subsection{Third specification}
The third specification moved linkability and attribute inference to a second tier, scored only on the pooled table and on groups of adequate size. We froze it and ran it once on five splits it had never seen. Clean passed 0 of 5 at tolerances of 0.02, 0.03 and 0.05. Three causes appeared.

\paragraph{An absolute coverage floor.} The floor was 0.80 and clean coverage measured 0.300 to 0.333. The clean release was half the size of the holdout. A holdout row counts as covered when its nearest release neighbour is at least as close as its nearest other holdout row. If release and holdout are exchangeable samples from one distribution, the nearest of the $m+h-1$ other points is a release point with probability
\begin{equation}
\mathbb{E}[K_1]=\frac{m}{m+h-1},
\end{equation}
which is close to one third for $m=4{,}521$ and $h=9{,}043$. The floor was unreachable by construction. Coverage needs a size-matched reference value, not a constant.

\paragraph{Residual groups.} The rule for merging small groups left groups of one or two records. Those groups have no valid decision.

\paragraph{Multiple testing.} Each release carried 52 to 102 mandatory components, and each clean split showed 0 to 2 exceedances. For example, density ratio on the vulnerable group in Diabetes returned an upper leak of 0.067 on 560 members. A component that passes clean data 95\% of the time gives no joint guarantee. Under independence, 52 such components all pass with probability $0.95^{52}\approx 7\%$; for 102 components the figure is about 0.5\%. Components are correlated, so the true figures are higher. A later check with the independent implementation (54 components, 50 clean Bank Marketing releases) shows that the problem depends on how a gate reads its components. Under our rule, which passes a component only when its upper bound is within tolerance, 46 of 50 clean releases passed at $t=0.05$ with or without a Holm correction, and 2 of 50 passed at $t=0.02$; requiring every component to pass already controls the risk of passing a leaky release, and correction only makes the rule stricter. A rule that fails a release on any evidence of a leak passed 13 of 50 uncorrected and 47 of 50 with Holm. The $0.95^{52}$ arithmetic describes the second kind of rule. For ours, the failures at tight tolerances come from small groups near the adequacy floor, and two components with fewer than 100 holdout records blocked every release.

With coverage and usefulness set aside, the privacy and copy checks alone passed the clean control on 2 of 5, 0 of 5 and 2 of 5 splits at the three tolerances.

\subsection{Harness errors}
The third round logged three defects in the test harness: a helper looked up the wrong frozen configuration, a loop variable shadowed its accessor, and an undersized copy component was reported as passed when it had no hits. All five splits were rerun from scratch after the fixes, and an audit recomputed all 6,945 component decisions from the raw record arrays.

\section{What this means for anyone releasing synthetic tables}\label{sec:guidance}
\begin{itemize}
\item \textbf{Compare every group share against the tolerance.} A pooled score says nothing about a group smaller than the tolerance. With a 5\% tolerance, list every class and segment under 5\% of rows and treat each as untested.
\item \textbf{Count before testing a group.} Check group size against the floors in Section~\ref{sec:floors}. Below 49 members at a 5\% tolerance, report the group as not assessed, never as passed.
\item \textbf{Run several attacks.} Pooled density ratio and pooled linkability missed an exposed class on 5 of 5 splits at every tolerance, while pooled distance caught it at 0.02 and 0.03. Our control favours a distance attack, so read this as a warning against relying on one attack, not as a ranking.
\item \textbf{Plant a clean control.} Most evaluations test whether a check catches a leak; few test whether it passes clean data. Ours failed the second test three times. A check nobody has run on clean data has an unknown false-alarm rate.
\item \textbf{Match coverage to release size.} A fixed coverage floor penalises small releases. Compare against $m/(m+h-1)$, the coverage an equally good sample of the same size achieves.
\item \textbf{Declare which error is controlled.} A gate with 52 to 102 components must say whether it controls the chance of passing a leaky release or of blocking a clean one. A rule that fails on evidence of a leak needs a family-wise correction such as Holm; a rule that passes only on evidence of safety needs groups large enough to clear the adequacy floor instead.
\item \textbf{Word claims to match the test.} An attack-based gate supports one statement: these attacks found no leak above tolerance $t$ in groups of at least $n$ records. It does not support a statement of the form ``no real person is exposed''.
\end{itemize}

\paragraph{An open problem.} A working per-group gate needs four things we did not deliver: a hierarchical or corrected decision rule across components, a size-aware coverage reference, a principled treatment of groups below the floor, and a confirmatory run on data no specification has touched. Adult and Default of Credit Card Clients remain unused for this purpose.

\section{Limitations}
\begin{enumerate}
\item \textbf{Gate results use planted controls only.} The three specifications were tested on planted controls. Trained generators were run only for the masking result, on one table, three seeds and two attacks, with short training (30 CTGAN and 150 TVAE epochs) and an MST-style generator of our own rather than a reference implementation.
\item \textbf{Two tables.} Both are binary-class tabular datasets, and the rare class was created by downsampling to 3\%, not found in the wild.
\item \textbf{A small attack set.} We ran copy checks, distance, density ratio, linkability and attribute inference, and no shadow-model attack. Stronger attacks would change scores and leave the floors in place.
\item \textbf{The second round is retrospective.} Its amendments were written after seeing first-round data. Only the third round is prospective.
\item \textbf{The multiplicity figures in Section~\ref{sec:failures} assume independence.} The 50-release check covers one table and two attacks.
\item \textbf{One operator for the main harness.} A single AI coding agent implemented the harness and ran every gate test, and three harness defects surfaced in the third round. Only the masking result, the floors and the multiple-testing check have been reproduced independently (Appendix~\ref{app:replication}).
\item \textbf{No systematic literature review.} Section 2 covers the work we found, not the field.
\item \textbf{A limited look at differential privacy.} One MST-style generator at $\varepsilon=1$ and $8$ showed no rare-class leak to our attacks. A formal guarantee addresses rare-group exposure by a different route, and we did not audit it.
\end{enumerate}

\section{Reproducibility record and disclosure}\label{sec:disclosure}
\paragraph{Data.} Bank Marketing \citep{moro2014bank} and Diabetes 130-US Hospitals \citep{strack2014diabetes}, both from the UCI Machine Learning Repository. German Credit and SECOM entered data preparation and were excluded for group size.

\paragraph{Splits.} Bank Marketing seeds 1729, 1733, 1741 (rounds one and two) and 1747, 1753 (round three); Diabetes seeds 1729, 1733, 1741 (round three). Split manifest SHA-256: \hash{a18fcd23d88186475424031189d24354963466a09d10b12dc8a62835f29cf6ce}.

\paragraph{Frozen specifications.}
\begin{itemize}
\item First round. Freeze SHA-256 \hash{abd428e5109a037408150e0ceb71cf058d59aa98068808284bff037aec58a323}; run commit \hash{a8b0b69856164c4b1181e6661aa9825d65576273}.
\item Second round (replay, no freeze). Commit \hash{cca8b5205e4004b193d34f29b33a11dfa59394a7}.
\item Third round. Freeze SHA-256 \hash{50a7f2d568441286d53e2f4b349fe81814a098a5108361c6b8f2284eb8e51ba9}; protocol commit \hash{e3835fd}; run commit \hash{86caf24eee7c6aa437232f61dddc5976d9ea87ed}.
\end{itemize}

\paragraph{Code and logs.} The gate harness, run logs and component decisions sit in a private repository at the time of writing. The independent reimplementation of the masking result and the floor calculations is described in Appendix~\ref{app:replication}.

\paragraph{AI assistance.} AI assistants drafted the gate specifications and this text under the author's direction. A separate AI coding agent implemented the harness and ran the tests; the drafting assistant worked from the agent's written reports and did not inspect the code or the raw logs. Several specification errors reported in Section~\ref{sec:failures} originated with the drafting assistant. The author directed the work and is responsible for the content.

\paragraph{Conflict of interest.} The author is the founder of Fahrenheit Research, which is developing Grafta, a synthetic-table product with a release gate.

\appendix
\section{Independent reimplementation}\label{app:replication}
To test the masking rule without the private harness, we wrote a separate, minimal implementation from the paper's definitions alone. It shares no code with the gate harness. It uses the UCI Bank Marketing table, the same four-way split and 3\% rare-class arm, a Clean release drawn from the curation oracle, and a Rare-leak release formed by adding every rare-class training record, unchanged, to the Clean release. Numeric columns are standardised and categorical columns one-hot encoded with statistics from the reference split. The distance attack scores each record by the negative distance to its nearest release record, with the threshold set at the 99th percentile of holdout scores. Splits use NumPy's \texttt{default\_rng} with the paper's seeds; because the original generator differs, individual rows are not the same records as in the main study.

\begin{table}[h]
\centering\footnotesize\setlength{\tabcolsep}{4pt}
\caption{Independent reimplementation on Bank Marketing (distance attack, $\alpha=0.01$). Bound is the one-sided 95\% Clopper-Pearson upper bound on the pooled leak; $P_t$ is the pooled distance score at tolerance $t$.}\label{tab:replication}
\begin{tabular}{rrrcccccc}
\toprule
Seed & Train & Rare & Recall & Leak $u$ & Bound & $P_{0.02}$ & $P_{0.05}$ & Clean $P_{0.05}$\\
\midrule
1729 & 24,694 & 741 & 741/741 & 0.0326 & 0.0348 & 75.52 & 90.21 & 99.30\\
1733 & 24,723 & 742 & 742/742 & 0.0339 & 0.0361 & 74.57 & 89.83 & 98.76\\
1741 & 24,727 & 742 & 742/742 & 0.0304 & 0.0325 & 77.21 & 90.89 & 99.98\\
1747 & 24,702 & 741 & 741/741 & 0.0308 & 0.0330 & 76.88 & 90.75 & 99.83\\
1753 & 24,689 & 741 & 741/741 & 0.0320 & 0.0342 & 76.00 & 90.40 & 99.47\\
\bottomrule
\end{tabular}
\end{table}

Every seed recovered the entire rare class, failed the pooled check at $t=0.02$ and $t=0.03$, and passed it at $t=0.05$ with a score between 89.83 and 90.89, within 0.9 points of the main study. The measured leak sits 0.002 to 0.004 above $g$ because the realised false-positive rate on clean members ran between 1.01\% and 1.41\%, mostly because of class mix rather than noise. Training members in the 3\% arm are 3\% positive while the holdout is about 11.7\% positive, and positive records sit farther from the mostly negative release (median distance 2.15 against 1.50 on seed 1729). On clean releases the pooled distance component therefore raised an alarm in 32\% of 50 releases, against about 5\% for the same attack within each class. A sweep over group shares from 0.5\% to 10\%, using random subgroups fully exposed in the same way, crosses the pass mark at $g\approx t$ for all three tolerances (Figure~\ref{fig:masking}). The same code reproduces the detection floors (100, 74, 49) and adequacy floors (566, 296, 152) of Table~\ref{tab:floors} exactly. The same implementation reran the Diabetes 130-US splits (Table~\ref{tab:diabetes}). The distance attack again recovered every exposed record and the pooled check failed at $t=0.02$ and $0.03$ and passed at $0.05$ on every seed. The pooled density-ratio attack found 214 to 218 of about 1,678 exposed records and passed at every tolerance, matching Table~\ref{tab:pooled}.

\begin{table}[h]
\centering\footnotesize\setlength{\tabcolsep}{4pt}
\caption{Independent reimplementation on Diabetes 130-US Hospitals (distance attack, $\alpha=0.01$). Columns as in Table~\ref{tab:replication}.}\label{tab:diabetes}
\begin{tabular}{rrrcccccc}
\toprule
Seed & Train & Rare & Recall & Leak $u$ & Bound & $P_{0.02}$ & $P_{0.05}$ & Clean $P_{0.05}$\\
\midrule
1729 & 55,918 & 1,678 & 1,678/1,678 & 0.0323 & 0.0337 & 75.81 & 90.32 & 99.44\\
1733 & 55,771 & 1,673 & 1,673/1,673 & 0.0303 & 0.0317 & 77.30 & 90.92 & 100.00\\
1741 & 55,989 & 1,680 & 1,680/1,680 & 0.0322 & 0.0336 & 75.86 & 90.34 & 99.25\\
\bottomrule
\end{tabular}
\end{table}

The scripts, figures and result files accompany this paper as ancillary material.

\bibliographystyle{plainnat}
\bibliography{refs}
\end{document}